\documentstyle[12pt]{article}
 \newcommand{\up}{\vspace{-3mm}}
 
\begin{document}
 \begin{center}
 {\large\bf Quantum Interferometry: Some Basic Features Revisited}
 \\[6mm]
 {Markus Simonius}\\[2mm]
 Inst. f. Teilchenphysik, Eidg. Techn. Hochschule,
 CH-8093 Z\"urich, Switzerland\\[6mm]

 \parbox{14cm}{\small
 The reduction paradigm of quantum interferometry is reanalyzed. In
 contrast to widespread opinion it is shown to be amenable to
 straightforward mathematical treatment within ``every-users''
 simple-minded single particle quantum mechanics (without reduction
 postulate or the like), exploiting only its probabilistic content.
 }\vspace{3mm}\end{center}

 Consider a typical interferometer arrangement as sketched in 
 Fig.~1.
 Its properties are well known: even if only one quantum (photon or
 neutron etc.) is inside the arrangement at a given time, the
 configuration of Fig.~1a can reveal interference between the states
 passing the two arms I and II provided none of them is blocked.
 Correspondingly, a pure single-particle state within the interferometer is
 represented by a normed wave function of the form 
 \begin{equation} \label{e1}
 \varphi = c_1\varphi_1+c_2\varphi_2,\ \
 \|\varphi\|=\|\varphi_1\|=\|\varphi_2\|=1, 
 \end{equation}
 where, $\varphi_1$ and $\varphi_2$ represent states passing completely
 along one of the two paths I or II, respectively, in the
 interferometer and have zero component in the other (and thus are
 mutually orthogonal). For ideal 50:50 beam-splitting $|c_1|^2 =
 |c_2|^2 = {1\over 2}$, of course.

 \begin{figure}[h] 
 \vspace{-6mm}
 \center
 \begin{minipage}[c]{4.9cm} \center %Fig. 1a
 \setlength{\unitlength}{0.0054in}% mach-zehnder1a
 \begin{picture}(383,121)(45,659)
 \thicklines
 \put(398,769){\line( 1,-2){ 18}}
 \multiput(416,733)(1.20000,0.60000){11}{\makebox(1.0336,1.5504){\sevrm .}}
 \put(428,739){\line(-1, 2){ 18}}
 \multiput(410,775)(-1.20000,-0.60000){11}{\makebox(1.0336,1.5504){\sevrm .}}
 \put(398,670){\line( 1, 2){ 18}}
 \multiput(416,706)(1.20000,-0.60000){11}{\makebox(1.0336,1.5504){\sevrm .}}
 \put(428,700){\line(-1,-2){ 18}}
 \multiput(410,664)(-1.20000,0.60000){11}{\makebox(1.0336,1.5504){\sevrm .}}
 \put( 45,690){\line( 2, 1){180}}
 \put(225,780){\line( 2,-1){120}}
 \put(105,720){\line( 2,-1){120}}
 \put(225,660){\line( 2, 1){180}}
 \multiput(345,720)(8.57143,-4.28571){8}{\makebox(1.0336,1.5504){\tenrm .}}
 \multiput( 45,690)(1.20000,0.60000){21}{\makebox(1.0336,1.5504){\sevrm .}}
 \put( 69,702){\vector( 2, 1){0}}
 \multiput( 72,720)(10.15385,0.00000){7}{\line( 1, 0){  5.077}}
 \multiput(312,720)(10.15385,0.00000){7}{\line( 1, 0){  5.077}}
 \put(195,659){\line( 1, 0){ 60}}
 \put(196,780){\line( 1, 0){ 60}}
 \put(156,756){\makebox(0,0)[lb]{\raisebox{0pt}[0pt][0pt]{\ninrm I}}}
 \put(153,666){\makebox(0,0)[lb]{\raisebox{0pt}[0pt][0pt]{\ninrm II}}}
 \end{picture}
 \\ \ a
 \end{minipage}   \hspace{6mm}
 \begin{minipage}[c]{4.9cm} \center %Fig. 1b
 \setlength{\unitlength}{0.0054in}%mach_zehnder1b
 \begin{picture}(360,121)(45,659)
 \thicklines
 \put( 45,690){\line( 2, 1){180}}
 \put(225,780){\line( 2,-1){120}}
 \put(105,720){\line( 2,-1){120}}
 \put(225,660){\line( 2, 1){180}}
 \put(195,780){\line( 1, 0){ 60}}
 \multiput(345,720)(8.57143,-4.28571){8}{\makebox(1.0336,1.5504){\tenrm .}}
 \multiput( 45,690)(1.20000,0.60000){21}{\makebox(1.0336,1.5504){\sevrm .}}
 \put( 69,702){\vector( 2, 1){0}}
 \multiput( 72,720)(10.15385,0.00000){7}{\line( 1, 0){  5.077}}
 \multiput(312,720)(10.15385,0.00000){7}{\line( 1, 0){  5.077}}
 \put(267,704){\line( 1,-2){ 18}}
 \multiput(285,668)(1.20000,0.60000){11}{\makebox(1.0336,1.5504){\sevrm .}}
 \put(297,674){\line(-1, 2){ 18}}
 \multiput(279,710)(-1.20000,-0.60000){11}{\makebox(1.0336,1.5504){\sevrm .}}
 \put(267,737){\line( 1, 2){ 18}}
 \multiput(285,773)(1.20000,-0.60000){11}{\makebox(1.0336,1.5504){\sevrm .}}
 \put(297,767){\line(-1,-2){ 18}}
 \multiput(279,731)(-1.20000,0.60000){11}{\makebox(1.0336,1.5504){\sevrm .}}
 \put(196,659){\line( 1, 0){ 60}}
 \put(156,756){\makebox(0,0)[lb]{\raisebox{0pt}[0pt][0pt]{\ninrm I}}}
 \put(153,666){\makebox(0,0)[lb]{\raisebox{0pt}[0pt][0pt]{\ninrm II}}}
 \end{picture}
 \\ \ b
 \end{minipage}   \hspace{3mm}
 \begin{minipage}[c]{4.9cm} \center %Fig. 1c
 \setlength{\unitlength}{0.0054in}%mach_zehnder1c
 \begin{picture}(360,121)(45,659)
 \thicklines
 \put( 45,690){\line( 2, 1){180}}
 \put(225,780){\line( 2,-1){120}}
 \put(105,720){\line( 2,-1){120}}
 \put(225,660){\line( 2, 1){180}}
 \put(195,780){\line( 1, 0){ 60}}
 \multiput(345,720)(8.57143,-4.28571){8}{\makebox(1.0336,1.5504){\tenrm .}}
 \multiput( 45,690)(1.20000,0.60000){21}{\makebox(1.0336,1.5504){\sevrm .}}
 \put( 69,702){\vector( 2, 1){0}}
 \multiput( 72,720)(10.15385,0.00000){7}{\line( 1, 0){  5.077}}
 \multiput(312,720)(10.15385,0.00000){7}{\line( 1, 0){  5.077}}
 \put(257,740){\line( 1, 2){ 18}}
 \multiput(275,776)(1.20000,-0.60000){11}{\makebox(1.0336,1.5504){\sevrm .}}
 \put(287,770){\line(-1,-2){ 18}}
 \multiput(269,734)(-1.20000,0.60000){11}{\makebox(1.0336,1.5504){\sevrm .}}
 \put(281,729){\line( 1, 2){ 18}}
 \multiput(299,765)(1.20000,-0.60000){11}{\makebox(1.0336,1.5504){\sevrm .}}
 \put(311,759){\line(-1,-2){ 18}}
 \multiput(293,723)(-1.20000,0.60000){11}{\makebox(1.0336,1.5504){\sevrm .}}
 \put(196,659){\line( 1, 0){ 60}}
 \put(156,756){\makebox(0,0)[lb]{\raisebox{0pt}[0pt][0pt]{\ninrm I}}}
 \put(153,666){\makebox(0,0)[lb]{\raisebox{0pt}[0pt][0pt]{\ninrm II}}}
 \end{picture}
 \\ \ c
 \end{minipage}
 \caption{Sketch of typical interferometer arrangements with
 differently placed detectors.}\label{fig1}
 \end{figure}
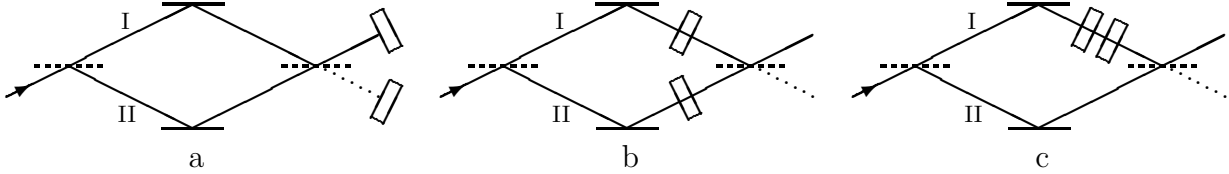

 Now consider instead coincidences between two detectors inserted into
 the two paths of the interferometer as shown in Fig.\ \ref{fig1}b.
 From the appearance of the wave function (\ref{e1}) one might suspect
 that such coincidences should occur. On the other hand, it is the
 fundamental proposition of quantum physics that quanta (particles or
 photons) are {\em indivisible entities}
 and thus a {\em single quantum} is not able to
 trigger both detectors in the arrangement of Fig.\ \ref{fig1}b (nor in
 Fig.\ \ref{fig1}a, of course).
 Though all this is much discussed textbook wisdom (tested also
 experimentally \cite{Aspect}), one looks in vain in the
 literature for a satisfying {\em derivation} of this fact which uses
 only the basic structure of single particle quantum theory and does
 not amount to just stating the fact as postulate in one way or
 another.
 But clearly it should be possible to deduce such
 basic properties mathematically once the general premises of the theory are
 laid down \cite{Bell90a}. It is the object of this note to show that
 this is indeed so. Astonishingly, though the analysis to be presented
 is simple and straightforward, no such treatment is found in the
 literature, let alone in textbooks where it would belong.

 For simplicity only pure states will be considered here explicitly,
 but the same results are obtained also directly from the general
 definition of superpositions in Ref. \cite{Simonius93} which applies
 also to non-pure states (density operators).

 {\em Only the following minimal probabilistic set of postulates of
 quantum mechanics is used. No
 state reduction and no axiom of measurement etc.!}

 \begin{list}{}{
 \topsep3pt
 \itemsep0pt plus 1pt minus 1pt
 \parsep0pt plus 1pt minus 1pt} \em
 \item[I ] Pure states are represented by normed elements $\varphi \in
 {\cal H}$ of a Hilbert space ${\cal H}$.
 \item[II] The probability for a given kind of event for a system in a
 state represented by $\varphi$ is given by an expectation value
 $\langle\varphi|A|\varphi\rangle$ where $A$ is a hermitian operator on
 ${\cal H}$ which obviously must obey
 $0\le\langle\varphi|A|\varphi\rangle \le 1$ for all $\varphi\in {\cal
 H},\ \|\varphi\|=1$.
 \end{list}
 The positivity postulate in II immediately leads to the
 following ``silly'' but far reaching

 \noindent{\bf Theorem:} {\em 
 Let $A$ be a positive operator on ${\cal H}$ such that
 $\langle\psi_1|A|\psi_1\rangle =
 \langle\psi_2|A|\psi_2\rangle =0$ for given
 $\psi_1,\psi_2\in{\cal H}$.  Then
 $\langle\psi|A|\psi\rangle = 0$ for all superpositions
 $\psi = c_1\psi_1+c_2\psi_2$ between them}.

 In order to apply this theorem to Fig.\ \ref{fig1}b with the two detectors
 in coincidence one only has to remark that one of the fundamental
 requirements for a coincidence between the two detectors is that the
 probability for a coincidence event be zero for any state which has
 zero component in one of the two arms of the interferometer, i.e if
 either $c_1=0$ or $c_2$=0 in eq.\ (\ref{e1}).  (This is actually what
 careful experimenters check in order to verify that their arrangement
 does not produce spurious coincidence events!)  Thus
 $\langle\varphi_1|A|\varphi_1\rangle = 
 \langle\varphi_2|A|\varphi_2\rangle =0$ 
 where $A$ is the operator describing the probability of coincidence
 events. It then follows from the above theorem that
 $\langle\varphi|A|\varphi\rangle = 0$ also for arbitrary $\varphi =
 c_1\varphi_1+c_2\varphi_2$ and thus that {\em also for arbitrary
 superpositions between the states in the two arms of the
 interferometer the two detectors in Fig.\ \ref{fig1}b have zero
 probability to produce a coincidence event.}

 Thus ``what must be'' is graciously born out by the mathematical
 analysis in spite of appearance of the wave function in eq.\
 (\ref{e1}), ``with nothing left to the discretion of the theoretical
 physicist'' \cite{Bell90a} (except to formulate the problems properly).

 A corresponding ``reduction theorem'' is obtained similarly (using
 anticoincidence, this time) for an arrangement
 of the kind shown in Fig. 1c where now both detectors are in
 the same path and it is assumed that at least the first detector
 transmits (does not absorb) the quanta: If the
 two detectors have unit efficiency either both of them fire or none.

 It is emphasized that postulate II is an integral part of the
 mathematical structure of quantum theory and not just a supplementary
 interpretation. In fact this postulate distinguishes quantum theory
 from classical wave theory.

 The structure of the ``silly'' theorem on which this is based and of
 the corresponding physical statements is emphasized: Conclusions for
 arbitrary superpositions between two states $\varphi_i$ are obtained
 from conditions imposed only for the two states $\varphi_i$
 themselves.  There is no need to {\em postulate} what happens in the
 case of superpositions. The results presented depend exclusively on
 properties of the Hilbert space used to describe an elementary system
 and the operators acting on it.  Indeed, in the case analyzed
 explicitly here, it suffices to take into account the two-dimensional
 space spanned by $\varphi_1$ and $\varphi_2$ in eq.\ (\ref{e1}) which
 supports only four linearly independent operators.

 Two-slit or Stern-Gerlach arrangements can be
 analyzed correspondingly.

 Of course, the features proved here must be revealed also if the
 detection devices or whatever are taken into account explicitly.
 However, against widespread opinion \cite{Gottfried91a}, it is
 definitely not {\em necessary} to do so let alone to invoke
 corresponding macroscopic properties.

 \end{document}